\documentclass[twocolumn,aps,showpacs,prb,tightenlines,amsmath,amssymb,superscriptaddress]{revtex4}
\usepackage{graphicx}
\usepackage{amssymb}
\usepackage{dcolumn} 
\usepackage{amsmath}
\usepackage{bm}
\usepackage{colordvi}
\newcommand{\bgreek}[1]{\mbox{\boldmath$#1$\unboldmath}}
\makeatletter

\newcommand{\Rmnum}[1]{\expandafter\@slowromancap\romannumeral #1@}
\makeatother

\begin{document}

\title{Spin relaxation in $n$-type (111) GaAs quantum wells}
\author{B. Y. Sun}
\affiliation{Hefei National Laboratory for Physical Sciences at
  Microscale,
University of Science and Technology of China, Hefei,
  Anhui, 230026, China}
\affiliation{Department of Physics,
University of Science and Technology of China, Hefei,
  Anhui, 230026, China}
\author{P. Zhang}
\affiliation{Department of Physics,
University of Science and Technology of China, Hefei,
  Anhui, 230026, China}
\author{M. W. Wu}
\thanks{Author to  whom correspondence should be addressed}
\email{mwwu@ustc.edu.cn.}
\affiliation{Hefei National Laboratory for Physical Sciences at
  Microscale,
University of Science and Technology of China, Hefei,
  Anhui, 230026, China}
\affiliation{Department of Physics,
University of Science and Technology of China, Hefei,
  Anhui, 230026, China}
\date{\today}

\begin{abstract}
We investigate the spin relaxation limited by the D'yakonov-Perel' mechanism in
$n$-type (111) GaAs quantum wells, by means of the kinetic spin Bloch equation
approach.  In (111) GaAs quantum wells, the in-plane 
effective magnetic field from
the D'yakonov-Perel' term can be suppressed to zero on a special 
momentum circle under the
proper gate voltage, by the cancellation between the Dresselhaus and
Rashba spin-orbit coupling terms. When the spin-polarized electrons mainly
distribute around this special circle,  the in-plane inhomogeneous
broadening is small and the spin relaxation can be suppressed,
especially for that along the growth direction
of quantum well. This cancellation
effect may cause a peak (the cancellation peak) in the 
density or temperature dependence of
the spin relaxation time. In the density (temperature) dependence,
the interplay between the  cancellation peak and the
ordinary density (Coulomb) peak leads to rich features of the  
density (temperature) dependence of the spin relaxation time.
The effect of impurities, with its different weights on the cancellation peak 
and the Coulomb peak in the temperature dependence of the spin 
relaxation, is revealed.
We also show the anisotropy of the spin relaxation with respect to
the spin-polarization direction. 
\end{abstract}

\pacs{72.25.Rb, 73.21.Fg, 71.10.-w, 71.70.Ej}
\maketitle
\section{INTRODUCTION}
In the past decades, great efforts have been devoted to the understanding of the
spin relaxation in various systems, aiming to incorporate the spin degree
of freedom of carriers into the traditional electronic
devices.\cite{opticalorientation,Wolf,Chen,wuReview,DasSarmaetal,Korn}
In $n$-type III-V zinc-blende semiconductors, the spin relaxation is mainly
governed by the D'yakonov-Perel' (DP) mechanism\cite{Dyakonov} via the joint
effects of the momentum scattering and the ${\bf k}$-dependent effective magnetic
field (inhomogeneous broadening\cite{wu3}) induced by the DP
spin-orbit coupling.\cite{wuReview} The DP spin-orbit coupling is contributed by the
Dresselhaus\cite{Dresselhaus} term as well as the possible Rashba\cite{Rashba}
term if the structure inversion symmetry is broken. In quantum-well structure,
the effective magnetic field
from the gate-voltage tunable Rashba term is proportional to ${\bf
  k}\times{\bf \hat{z}}$ (${\bf \hat{z}}$ is the growth direction), whereas that from the
Dresselhaus term varies with the growth direction. Therefore, with special growth direction,
spin-polarization direction and/or relative strength of the Rashba and
Dresselhaus terms, spin relaxation may show intriguing properties. In
(001) asymmetric GaAs quantum wells, the relaxation for spins along
[110] or [1${\bar 1}0$] can be strongly suppressed when the Rashba and Dresselhaus terms are
comparable in magnitude.\cite{Averkiev} In (110) symmetric GaAs quantum wells, the effective
magnetic field solely contributed by the Dresselhaus term is along the
growth direction.\cite{Winkler} Therefore for spins along the growth direction
the DP relaxation mechanism is absent and other relaxation
mechanisms should be taken into account.\cite{Ohno,wu2,Sherman110,Sherman,Dohrmann,Muller,YZhou1,YZhou2}

For (111) III-V zinc-blende quantum wells, the spin 
relaxation may also show rich properties, due to the
interplay of the Rashba and Dresselhaus terms. Setting the
${\bf \hat{z}}$-axis along the growth direction [111], ${\bf \hat{x}}$-axis along
[1$\bar{1}$0] and ${\bf \hat{y}}$-axis along [11$\bar{2}$], the DP term for the
lowest subband can be expressed as
\begin{eqnarray}
&&\Omega_{x}({\bf k})=\gamma(\frac{-k^2+4{\langle k_z^2\rangle}_{00}}{2\sqrt{3}})k_y-\alpha
      eE_zk_y,  \label{eq1}\\
&&\Omega_{y}({\bf k})=-\gamma(\frac{-k^2+4{\langle k_z^2\rangle}_{00}}{2\sqrt{3}})k_x+\alpha
      eE_zk_x,  \label{eq2}\\
&&\Omega_{z}({\bf k})=\gamma\frac{k_x^3-3k_xk_y^2}{\sqrt{6}}.\label{eq3}
\end{eqnarray}
Here terms containing coefficient $\gamma$ ($\alpha$) are contributed
by the Dresselhaus (Rashba) spin-orbit coupling; $E_z$ stands for the electric
field from the gate voltage and ${\langle k_z^2\rangle}_{00}=(\pi/a)^2$ is the average of the operator
$-(\partial/\partial z)^2$ over the electron state of the lowest
subband under the infinite-depth square well assumption. ${\bf \Omega}({\bf k})$ is
 referred to as the effective magnetic field, while the
spin-orbit coupling term in the Hamiltonian introduced by the DP mechanism is
expressed as $H_{so}={\bf \Omega}({\bf k})\cdot{\bgreek \sigma}/2$. Based on
Eqs.\,(\ref{eq1})-(\ref{eq3}), Cartoix\`a {\em et al.} proposed
that a peak of the spin relaxation time (SRT) in the gate-voltage dependence appears
at $E_z\approx 4{\langle k^2_z \rangle}_{00}\gamma/(2\sqrt{3}\alpha e)$ in (111) GaAs
quantum wells when the cubic term in ${\bf \Omega}({\bf k})$ is
neglected.\cite{Cartoixa} In (111) InGaAs quantum wells, Vurgaftman and Meyer also
investigated the spin relaxation, showing that the temperature
affects the gate-voltage dependence of the inhomogeneous broadening strongly and hence the SRT.\cite{Vurgaftman} Both investigations are based on the
single-particle approach.

\begin{figure}[htb]
\includegraphics[width=8.5cm]{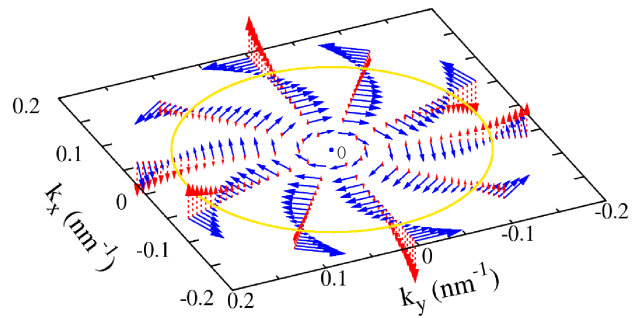}
\caption{(Color online) Schematic of the effective magnetic field
  ${\bf\Omega}({\bf k})$ in momentum
  space. The blue solid arrows represent the in-plane components of the
  effective magnetic field and the red dashed ones correspond to the
  out-of-plane ones. The point zero stands for the
  origin of the {\bf k} space. The in-plane effective magnetic field is zero on
  the  yellow circle, where $k=k_c$
  [Eq.\,(\ref{cancellationKc})]. In this schematic $E_z=80$~kV/cm.}
\label{figszw1}
\end{figure}

Besides the gate-voltage dependence, the electron density and temperature dependences of the SRT can also
show special properties. For convenience, we rewrite the in-plane DP term as
\begin{equation}
{\bf \Omega}_{\perp}({\bf k})=[(k^2-4{\langle
    k_z^2\rangle}_{00})\gamma/2\sqrt{3}+\alpha eE_z]{\bf \hat{z}}\times{\bf k}.
\label{inplanedp}
\end{equation}
From this equation, one finds that when $k^2$ is equal to the critical value modulated by the gate
voltage:
\begin{equation}
  k^2_c\equiv4{\langle k^2_z \rangle}_{00}-2\sqrt{3}\alpha e E_z/\gamma,
\label{cancellationKc}
\end{equation}
 ${\bf \Omega}_{\perp}({\bf k})$ becomes zero as
shown by the circle in the schematic of ${\bf\Omega}({\bf k})$
in Fig.~\ref{figszw1}. It is noted that this phenomenon only happens when $E_z<4{\langle
  k^2_z \rangle}_{00}\gamma/(2\sqrt{3}\alpha e)$, i.e., $k_c^2>0$.
Under this condition, the SRT of spins polarized
 along the ${\bf \hat{z}}$-axis tends to infinity if the
spin-polarized electrons distribute exactly on the critical
circle in the limit of zero temperature. However, in
  reality, the temperature is higher than zero and the
distribution of the spin-polarized electrons spreads
out around the average Fermi momentum. Also, the electron density can be
changed and hence the average Fermi momentum can be shifted away
  from $k_c$. Even so, with proper temperature and/or electron
  density, the condition $\langle k^2\rangle=k_c^2$ can be satisfied and
  then the in-plane inhomogeneous broadening $\langle{\bf \Omega}^2_{\perp}({\bf
  k})\rangle$ is suppressed. This may lead to the nonmonotonicity of
  the inhomogeneous broadening in the temperature or electron
  density dependence. In the following, we refer to this effect as
  the ``cancellation effect'' due to the cancellation between the Dresselhaus
and Rashba terms. It is noted that the cancellation
  effect is more obvious at low temperature. It
 may also take place when considering spin relaxations along
other directions. However, when the spin polarization  is
deviated from the ${\bf \hat{z}}$-axis, the cancellation effect
becomes weaker as the inhomogeneous broadening from $\Omega_z({\bf k})$ comes
  into play. The nonmonotonicity of the inhomogeneous broadening
may lead to a peak of the SRT in the temperature or electron
  density dependence. We call this peak as the 
``cancellation peak'', which to our knowledge has not been studied in the
literature and is hence the main focus of this work.

However, the scenario is not as simple as that presented above, since the spin relaxation is determined by the
joint effects of the inhomogeneous broadening
and the scattering. The scattering together with the inhomogeneous
broadening further lead to a plenty of features in the electron density or
temperature dependence of the SRT. In fact, as revealed in the previous works, the Coulomb scattering plays an
important role in the spin relaxation.\cite{wuReview,wu3,Glazov,weng2,leyland}
The nonmonotonic dependence of the Coulomb scattering
rate $1/\tau_{ee}$ on temperature $T$ during the crossover of electrons from the degenerate to
nondegenerate limits [e.g., for two-dimensional electrons 
$1/\tau_{ee}\propto$ $T^2$ ($T^{-1}$) when  $T\ll T_F$ ($T\gg T_F$)
 with $T_F$ denoting the Fermi temperature\cite{Giuliani}] can lead to
a peak in the temperature dependence of the
SRT in the strong scattering limit when the Coulomb scattering
dominates.\cite{Zhou,Bronold,Ruan,Zhang,Jiangbulk} This peak is called
as the ``Coulomb peak'' in the temperature dependence. It appears
  around $T_F$, with the location influenced by the temperature
  dependence of the inhomogeneous broadening and thus being sample
  dependent.  The increase of the inhomogeneous
broadening with increasing temperature tends to shift the peak towards
    a lower temperature. The peak was predicted to be close to $T_F$
  in $n$-type high-mobility (001) GaAs quantum wells\cite{Zhou} while later
  observed experimentally at about $T_F/2$ in $n$-type
  high-mobility (001) GaAs/GaAlAs heterostructure.\cite{Ruan} In $p$-type high-mobility (001) Si/SiGe (Ge/SiGe)
  quantum wells, the peak was predicted to be at about $T_F$
  ($T_F/2$).\cite{Zhang}  In intrinsic bulk
GaAs, this peak was predicted to be in the range of ($T_F/4$,
 $T_F/2$).\cite{Jiangbulk}
 Nevertheless, when impurities are added, the Coulomb peak 
can be destroyed.\cite{Zhang,Zhou,Jiangbulk} 
For the carrier density dependence of the
SRT, a peak also appears around the crossover of the nondegenerate and degenerate limits.\cite{Jiangbulk,Zhang,Kraub,shen}
This is because the inhomogeneous broadening varies little with density
in the nondegenerate limit where the Boltzmann distribution can
  be well satisfied but increases with increasing density in the
degenerate limit [this monotonic dependence widely exists in the
systems where the cancellation effect is absent, such as (001) GaAs quantum
wells or bulk GaAs]. Therefore, if the total scattering rate increases
with increasing density in the nondegenerate limit, the SRT increases
also provided the system is in the strong scattering limit. In the degenerate
limit, the increase of the inhomogeneous broadening with increasing density is
faster than that of the scattering, and thus the SRT decreases with
the increase of density. Consequently a peak (the normal density peak)
appears.\cite{Jiangbulk,Zhang,Kraub,shen,ma,jiangcomment} It is
 noted that all the scatterings can contribute to this peak. 
Nevertheless, in the system where the Coulomb scattering dominates,
this density peak is referred to as the Coulomb peak in the
 density dependence in this manuscript. Similar to the
  Coulomb peak in the temperature dependence, the location of the density
  peak is also sample dependent, usually with the corresponding Fermi
  temperature $T_F\sim$ $T/2$-$T$.\cite{Jiangbulk,Zhang,Kraub,shen,ma,jiangcomment} The above phenomena may also arise
here in (111) GaAs quantum wells and lead to a distinguishable
peak in the regime away from the cancellation peak. Moreover, when this peak and
the cancellation peak appear simultaneously, they may even interplay with each other.

In this work, we adopt the fully microscopic many-body
kinetic spin Bloch equation
(KSBE) approach\cite{wuReview} to investigate the spin relaxation
in (111) GaAs quantum wells, with all the relevant scatterings included. The electron density and temperature
dependences of the spin relaxation along the ${\bf \hat{z}}$-axis under different conditions (e.g.,
distinct gate voltages and impurity densities, etc.) are studied. Two peaks can be observed in the electron density
  dependence of the SRT when the
  temperature is low under proper gate voltages. One is the
  Coulomb peak while the other is the cancellation peak. The location
  of the former is insensitive to the gate voltage while that of the latter is modulated by
  the gate voltage. Under the proper gate voltage, these two peaks merge together
  and the SRT is largely prolonged. However, when the temperature
  increases, only one peak, the normal density peak, is observed. In
  the temperature dependence of the SRT, the Coulomb
  peak can be observed if the cancellation effect is excluded under a
  large gate voltage. When the cancellation effect is present under a relatively smaller gate
voltage, it together with the temperature dependence of the scattering lead to a single peak
in the temperature dependence of the SRT. The effect of
impurities on the temperature dependence of the spin relaxation is
investigated. Under the large gate voltage the Coulomb peak
is destroyed by the impurities. However, under the small gate
voltage where the cancellation effect is present, a peak always exists
even with very high impurity density. As the cancellation of
  the DP term only happens in the quantum-well plane, we also
  investigate the anisotropic spin relaxation by varying the
  spin-polarization direction. It is shown that when the
  spin-polarization  direction is tilted from the ${\bf \hat{z}}$-axis to
  the in-plane one, the cancellation peak gradually disappears due to
  the weakening of the cancellation effect.

This paper is organized as follows. In Sec. {\Rmnum 2}, we first
introduce the KSBEs and then investigate the spin relaxation in (111)
GaAs quantum wells by means of KSBEs. We summarize in Sec. {\Rmnum 3}.

\section{KSBEs AND NUMERICAL RESULTS}
We start our investigation from the $n$-type (111) GaAs quantum
wells. The well width $a$ is taken to be 7.5~nm. At this width including
only the lowest subband is sufficient in our investigation. The
 spin-polarization direction is along the ${\bf \hat{z}}$-direction except otherwise
  specified and the initial spin polarization is 0.025. The Rashba
parameter $\alpha$ is chosen to be 28~\AA$^2$.\cite{Hassenkam} The other
parameters ($\gamma$, effective electron mass and g-factor, etc.) are the
same as those in Ref.\,\onlinecite{weng}. The KSBEs read\cite{wuReview,wu3}
\begin{equation}
  \dot{\rho}_{\bf k}=\dot{\rho}_{\bf
    k}|_{\mbox{coh}}+\dot{\rho}_{\bf k}|_{\mbox{scat}}.
\end{equation}
Here $\rho_{\bf k}$ are the single-particle density matrices, whose
off-diagonal elements $\rho_{{\bf k}\frac{1}{2}-\frac{1}{2}}=\rho^{*}_{{\bf k}-\frac{1}{2}\frac{1}{2}}$
represent the spin coherence and the diagonal
elements $f_{{\bf k}\sigma}$ are electron distribution functions with spin
$\sigma$. Here $\sigma=1/2$ ($-1/2$) denotes the spin polarization along the
${\bf \hat{z}}$ ($-{\bf \hat{z}}$) axis. $\dot{\rho}_{\bf k}|_{\rm coh}$ are the coherent terms
describing the spin precession of electrons and $\dot{\rho}_{\bf k}|_{\rm
  scat}$ are the scattering terms including the electron-acoustic/longitudinal
optical phonon, electron-impurity and electron-electron Coulomb
scatterings. Their expressions are given in detail in Refs.~\onlinecite{Zhou}
and \onlinecite{weng}.

To numerically solve the KSBEs, the initial conditions are set as
\begin{equation}
  \rho_{\bf k}(t=0)= \frac{F_{{\bf k}\uparrow}+F_{{\bf
        k}\downarrow}}{2}+\frac{F_{{\bf k}\uparrow}-F_{{\bf
        k}\downarrow}}{2}{\bf \hat{n}}\cdot{\bgreek \sigma}.
\end{equation}
Here ${\bf \hat{n}}$ is the spin-polarization direction. $F_{{\bf k}\uparrow}$ ($F_{{\bf k}\downarrow}$) stand for the Fermi
distribution functions of electrons with spin-up (-down) determined by the polarized electron
density and temperature. By numerically solving the KSBEs, one can obtain
the single-particle density matrices at any time $t$. Then, the SRT can be obtained from the
temporal evolution of the spin polarization $P_{{\bf \hat{n}}}(t)=\sum_{\bf
  k}\mbox{Tr}[\rho_{\bf k}(t){\bf \hat{n}}\cdot{\bgreek
    \sigma}]/N_e$, with $N_e$ being the total electron density.

\begin{figure}[htb]
\includegraphics[width=7.5cm]{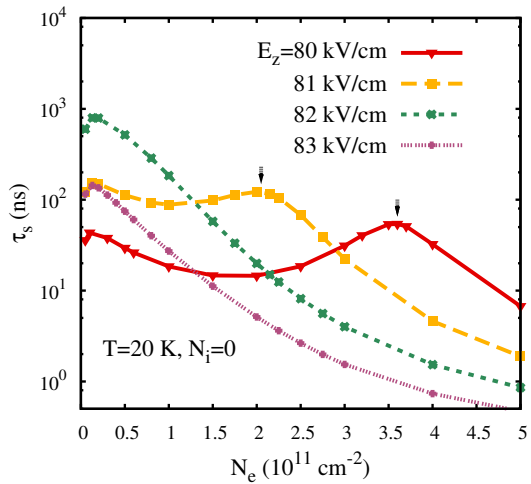}
\caption{(Color online) Electron density dependence of the SRT at different gate
  voltages. The temperature $T=20$~K and the impurity density $N_i=0$.}
\label{figszw2}
\end{figure}

\subsection{Electron density dependence of SRT}

We first study the electron density dependence of the SRT with different gate voltages at
$T=20$~K. The impurity density is zero. The results are plotted in Fig.~\ref{figszw2}.
From the figure, one finds that when $E_z<82$~kV/cm, two
peaks are present. When $E_z$ increases from 80 to 81~kV/cm, the peak
labeled with an arrow shifts towards the lower
density regime (from $N_e=3.6\times10^{11}$ to
$2.1\times10^{11}$~cm$^{-2}$), while the location of the
other peak remains unchanged at about
$N_e=0.2\times 10^{11}$~cm$^{-2}$. When $E_z\ge 82$~kV/cm, only one peak
can be observed at the location $N_e=0.2\times
10^{11}$~cm$^{-2}$. Moreover, the peak for the case with $E_z=82$~kV/cm
has a substantially longer SRT (up to $\sim$800~ns).

We first concentrate on the cases with $E_z< 82$~kV/cm where two peaks can be
observed. The peak, with its location remaining unchanged at
$\sim0.2\times10^{11}$~cm$^{-2}$ (the corresponding Fermi temperature for this density is
$T_F\approx 8$~K), is the Coulomb peak in the density dependence\cite{Jiangbulk,Zhang}
(here the Coulomb scattering is dominant). The location of the Coulomb peak is
insensitive to $E_z$, as the Coulomb scattering is
independent of $E_z$ and the variation rate of inhomogeneous
broadening versus $N_e$ is marginally affected by $E_z$ (since the
cancellation effect happens in the regime far away from the Coulomb
peak). The other peak (labeled with an arrow), with its location being tunable
by $E_z$, is just the cancellation peak. Based on
Eq.~(\ref{cancellationKc}), one can obtain the condition of the
gate voltage with which the cancellation effect can take place in the
density dependence. It is found that
$k_c^2<0$ when $E_z>E_z^c=82.8$~kV/cm. Therefore when $E_z>E_z^c$, $\langle{\bf \Omega}^2_{\perp}({\bf
  k})\rangle$ increases with $N_e$ monotonically and no cancellation
effect takes place. When $E_z<E_z^c$, with the increase of $E_z$,
$k_c^2$ decreases and thus the cancellation peak shifts 
towards the lower density regime. That is
exactly what is shown by the cases with $E_z< 82$~kV/cm in
Fig.~\ref{figszw2}. In addition, for the low temperature such as 20~K
here, Eq.~(\ref{cancellationKc}) can be utilized to estimate the location
of the cancellation peak by setting $k_F=k_c$ ($k_F$ is the Fermi momentum). The
corresponding electron densities $N^c_e$ obtained from $k_F$ are $3.8\times 10^{11}$ and $2.5\times 10^{11}$~cm$^{-2}$ for
the cases with $E_z=80$ and 81~kV/cm respectively. When $E_z=82$~kV/cm,
the two peaks merge together [the estimation from
  Eq.~(\ref{cancellationKc}) gives $N_e^c=1.1\times10^{11}$~cm$^{-2}$]
and the SRT is markedly prolonged. However, when $E_z=83$~kV/cm, the
cancellation effect is absent, causing the inhomogeneous broadening to
be stronger than the case with $E_z=82$~kV/cm
and to increase with $N_e$ monotonically. Therefore, the SRT with
$E_z=83$~kV/cm becomes smaller and the peak remaining at
$N_e=0.2\times10^{11}$~cm$^{-2}$ is again caused by the increase of
Coulomb scattering strength with increasing $N_e$ in the nondegenerate regime. In this sense, this peak can be
classified as the Coulomb peak.

\begin{figure}[htb]
\centering
\includegraphics[width=7cm]{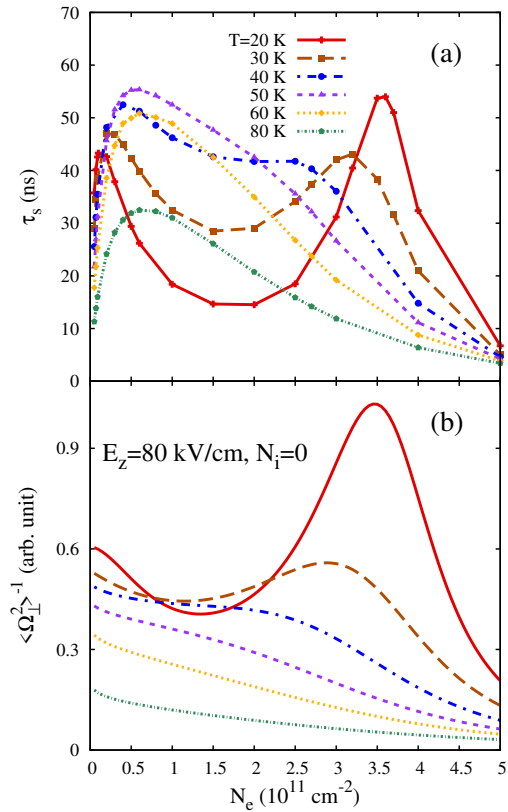}
\caption{(Color online) (a) Electron density dependence of the SRT at different
  temperatures. (b) The electron density dependence of $\langle
  {\bf\Omega}^2_{\perp}({\bf k})\rangle^{-1}$ corresponding to the cases shown
  in (a). Here $E_z=80$~kV/cm and the impurity density $N_i=0$.
}
\label{figszw3}
\end{figure}

We then investigate the electron density dependence of the SRT under different
temperatures with $E_z<E_z^c$, e.g., 80~kV/cm. The results are shown in
Fig.\,\ref{figszw3}(a). It is noted that with the increase of
$T$ from 20 to 50~K, the Coulomb (cancellation) peak gradually shifts 
towards the higher
(lower) density regime with increasing (decreasing) magnitude. Moreover, when $T\ge 50$~K, only one peak can
be observed, with the magnitude decreasing with the increase of
$T$. To facilitate the understanding of these phenomena, we further plot
the corresponding electron density dependences of the inverse of the inhomogeneous
broadenings, i.e., $\langle{\bf \Omega}^2_{\perp}({\bf
  k})\rangle^{-1}$, in Fig.\,\ref{figszw3}(b). The shift of the
Coulomb peak with $T$ from 20 to 50~K is understood by noticing
that the electron density corresponding to the crossover
of the nondegenerate and degenerate regimes increases with increasing
$T$ [it is noted that with increasing $N_e$, $1/\tau_{ee}$
increases in the nondegenerate regime
but decreases in the degenerate regime\cite{Giuliani}]. 
The increase in the magnitude of the Coulomb peak with increasing 
$T$ is due to the
enhancement of the Coulomb scattering (and the electron-phonon
scattering) as well as the small variation of the inhomogeneous
broadening with $T$ in the regime where the peak appears [as shown in
  Fig.\,\ref{figszw3}(b)]. With the increase of $T$, the cancellation 
peak decreases in magnitude
and shifts towards the lower density regime, as the corresponding 
peak in the $\langle{\bf \Omega}^2_{\perp}({\bf 
  k})\rangle^{-1}$-$N_e$ curve does [Fig.\,\ref{figszw3}(b)]. Finally
when $T$ exceeds 50~K, the cancellation peak is absent and the
single peak is the normal density peak rather than the Coulomb
  peak as the electron-phonon
scattering also comes into effect. The magnitude of this peak
decreases with increasing $T$ because the increase of the
inhomogeneous broadening suppresses the scattering strength.

\subsection{Temperature dependence of SRT}

\begin{figure}[htb]
\centering
\includegraphics[width=7cm]{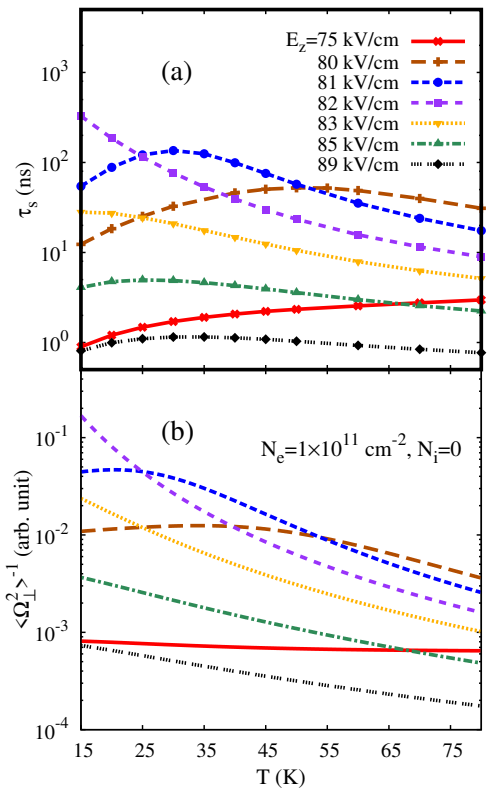}
\caption{(Color online) (a) The temperature dependence of the SRT
  in different gate voltages. (b) The
  corresponding temperature dependence of $\langle {\bf \Omega}^2_{\perp}({\bf
    k})\rangle^{-1}$. The electron density $N_e=10^{11}$~cm$^{-2}$
  and the impurity density $N_i=0$.}
\label{figszw4}
\end{figure}

We investigate the temperature dependence of the SRT in this
section. The study is first performed for high-mobility case under
different gate voltages. The electron density is taken to be $1\times10^{11}$~cm$^{-2}$ and the impurity
density is set to be zero. The results are plotted in
Fig.\,\ref{figszw4}(a). It is shown from the figure that when $E_z>
83$~kV/cm, a peak exists in the temperature dependence of the SRT and
shifts towards the lower temperature regime with the decrease of
$E_z$. When $E_z=82$~kV/cm, the SRT decreases with $T$
monotonically. However, as $E_z$ further decreases, a peak reappears
and shifts towards the higher temperature regime. Nevertheless, when
$E_z$ is as small as 75~kV/cm, the SRT increases with increasing $T$ monotonically and no peak is observed in the temperature regime under investigation.

To understand the features depicted above, we first specify at what
gate voltage the cancellation effect can take place in the temperature
dependence. From Eq.~(\ref{cancellationKc}), a critical
value of $E_z$ is obtained to be 82.1~kV/cm (referred to as $E_z^c$) by setting 
$k_c=k_F$. When $E_z>E_z^c$ ($<E_z^c$), $k_c^2<k_F^2$ ($>k_F^2$) and
the cancellation effect is absent (present). Moreover, when $E_z<E_z^c$, the smaller $E_z$ is, the
higher temperature at which the cancellation effect is expected to
happen becomes. However, due to the high temperature, the cancellation
effect becomes very weak. We further plot the
temperature dependence of $\langle{\bf \Omega}^2_{\perp}({\bf
k})\rangle^{-1}$ under different gate voltages in
Fig.~\ref{figszw4}(b). The properties of the $\langle{\bf
  \Omega}^2_{\perp}({\bf k})\rangle^{-1}$-$T$ curves are modulated by
the gate voltage in the way presented in the above theoretical analysis.

The peak observed in the $\tau_s$-$T$ 
curve with $E_z> 83$~kV/cm in Fig.~\ref{figszw4}(a) is actually the
Coulomb peak in the temperature dependence, as revealed previously in
other systems.\cite{Zhou,Bronold,Ruan,Zhang,Jiangbulk} Here the Fermi
temperature of the electrons is $T_F=40$~K, and the Coulomb peak is
located in the range of ($T_F/2$, $T_F)$. With the decrease of $E_z$,
the Coulomb peak shifts towards the lower temperature regime, and
finally disappears when $E_z$ approaches 82~kV/cm. This phenomenon is caused by
the increase in the decreasing rate of $\langle{\bf
  \Omega}^2_{\perp}({\bf k})\rangle^{-1}$ versus $T$, as shown in
Fig.~\ref{figszw4}(b). In fact, when $E_z=82$~kV/cm, which is almost equal to 
$E_z^c$, the in-plane DP term is substantially cancelled at $T=0$
and $\langle{\bf \Omega}^2_{\perp}({\bf k})\rangle^{-1}$ decreases
with increasing $T$ effectively. When
$E_z<82$~kV/cm, the cancellation effect arises and a peak of the SRT
reappears. The left-hand side of the peak is caused by both the increase of the 
scattering strength and the decrease of the inhomogeneous broadening
with increasing $T$, while the right-hand side caused by the increase of the
inhomogeneous broadening with increasing $T$. It is noted that the
location of this peak can be either lower or higher than $T_F$,
differing from the case of the Coulomb peak. In fact, this
peak is located at a temperature relatively higher than the
temperature corresponding to the peak in the $\langle{\bf
  \Omega}^2_{\perp}({\bf k})\rangle^{-1}$-$T$ curve, due to the
increase of the scattering strength with increasing $T$. With the
decrease of $E_z$, the peak of the SRT shifts towards the higher temperature
regime, as the cancellation effect takes place at a higher
temperature. Nevertheless, when $E_z$ is as small as 75~kV/cm, the SRT
increases with increasing $T$ monotonically. That is because at this
small gate voltage, the cancellation effect is expected to happen at
a high temperature beyond the regime under study. However, under this
high temperature the cancellation effect becomes very weak. As a result,
$\langle{\bf \Omega}^2_{\perp}({\bf k})\rangle^{-1}$ decreases with
increasing $T$ mildly in the temperature regime under study [as shown 
by the solid curve in Fig.~\ref{figszw4}(b)]. This mild decrease of $\langle{\bf
  \Omega}^2_{\perp}({\bf k})\rangle^{-1}$ is suppressed by the
increase of the scattering strength with increasing $T$ and consequently
the SRT increases with increasing $T$ monotonically. 

\begin{figure}[htb]
\centering
\includegraphics[width=7cm]{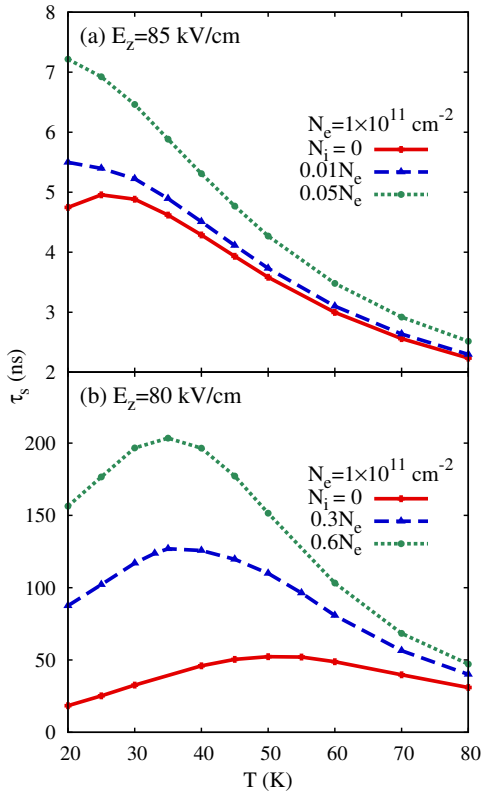}
\caption{(Color online) Temperature dependence of the SRT with different 
impurity densities. $N_e=1\times10^{11}$~cm$^{-2}$. (a) $E_z$=85~kV/cm
 and (b) $E_z$=80~kV/cm.}
\label{figszw5}
\end{figure}

We now investigate the effect of impurities on SRT under two typical
gate voltages, $E_z$=85 and 80~kV/cm, larger and smaller
than $E_z^c$ respectively. The temperature dependences of the
SRT with different impurity densities are plotted in
Fig.~\ref{figszw5}(a) and (b). In Fig.~\ref{figszw5}(a), the
peak disappears when even small amount of impurities are
present. However, in contrast, a peak always exists even at very high
impurity density but shifts towards the lower temperature regime with the increase of $N_i$ in Fig.~\ref{figszw5}(b). These phenomena are understood as follows. With the increase of $N_i$, the
electron-impurity scattering becomes dominant. However, the
electron-impurity scattering is insensitive to
$T$ when $T$ is low. Therefore, the temperature dependence of the
  SRT is determined by the variation of the inhomogeneous broadening
  with $T$. As a result, for the case with $E_z=85$~kV/cm where the
  inhomogeneous broadening increases with increasing $T$
  monotonically, the Coulomb peak is destroyed by the impurities as
  shown in Fig.~\ref{figszw5}(a). This scenario is the same as what
  predicted in (001) GaAs quantum wells.\cite{Zhou} However, for the
  case with $E_z=80$~kV/cm the peak always exists due to the
  cancellation effect. With the increase of $N_i$, this peak shifts towards
  the location corresponding to the peak in the $\langle{\bf \Omega}^2_{\perp}({\bf
    k})\rangle^{-1}$-$T$ curve.

\begin{figure}[bth]
\centering
\includegraphics[width=7cm]{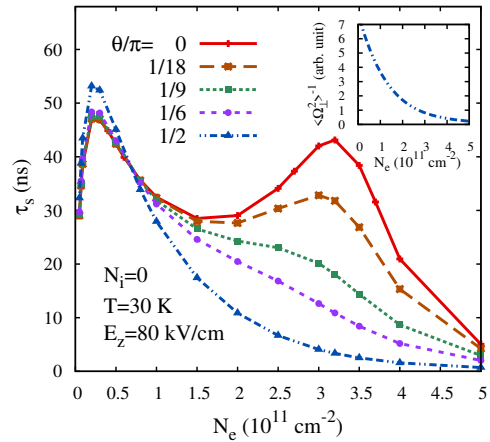}
\caption{(Color online) Electron density dependence of the SRT with different
  spin-polarization orientations. $\theta$ is
  the angle between the spin-polarization direction
and the ${\bf \hat{z}}$-axis. $T=30$~K, $E_z$=80~kV/cm and $N_i$=0. Inset shows
the electron density dependence of $\langle{\bf
  \Omega}^2_{\perp}({\bf k})\rangle^{-1}$ for the case with $\theta=\pi/2$.}
\label{figszw6}
\end{figure}

\subsection{Anisotropic spin relaxation}
As shown by Eqs.~(\ref{eq1})-(\ref{eq3}), the cancellation  of the
effective magnetic field only happens to the in-plane component. Therefore the
spin relaxation will show anisotropic property with the change of the polarization
direction. We vary the angle $\theta$ between the spin-polarization
direction and ${\bf \hat{z}}$-axis to investigate this anisotropy. The 
electron density dependence of the SRT with different values of
$\theta$ under temperature $T=30$~K is plotted in
Fig.~\ref{figszw6}. When $\theta=0$ the spins are polarized along the
${\bf \hat{z}}$-axis. The peak appears at $N_e\sim 3.2\times 10^{11}$~cm$^{-2}$ is the cancellation peak and
  that at $N_e\sim 0.2 \times 10^{11}$~cm$^{-2}$ is the Coulomb peak in
  density dependence, as discussed previously in Fig.~\ref{figszw3}(a). With the increase of $\theta$,
the cancellation peak gradually disappears. That is because the 
inhomogeneous broadening contributed by the ${\bf \hat{z}}$-component
of the effective magnetic field becomes more important and the
cancellation effect becomes weaker. In fact, when the
spin-polarization direction is in the ${\bf \hat{x}}$-${\bf \hat{y}}$
plane ($\theta=\pi/2$), e.g., along the ${\bf \hat{x}}$-axis, the inhomogeneous 
broadening, depicted by $\langle{\bf \Omega}^2_{\perp}({\bf
  k})\rangle=\langle\Omega^2_y({\bf k})+\Omega^2_z({\bf k})\rangle$,
increases monotonically with the increase of $N_e$, as shown in the inset of
Fig.\,\ref{figszw6}. Therefore, when $\theta=\pi/2$, only the Coulomb peak, but no cancellation peak, is
observed in the density dependence. The similar anisotropy of the spin relaxation also exists
in the temperature dependence. For
example, it is found that the peak in the dotted curve in
Fig.~\ref{figszw5}(b) gradually disappears when the spin-polarization
direction is tilted from the ${\bf \hat{z}}$-direction to
the in-plane one.

\section{CONCLUSION}

In conclusion, we have investigated the spin relaxation in $n$-type (111) GaAs
quantum wells by numerically solving the fully microscopic
KSBEs. Differing from the widely investigated (100) GaAs quantum
wells, in (111) GaAs quantum wells, the in-plane effective magnetic field from
the DP term can be suppressed to zero on a special momentum circle under the
proper gate voltage due to the cancellation of the Dresselhaus
and Rashba spin-orbit couplings. When the spin-polarized electrons mainly
distribute around this special circle,  the in-plane inhomogeneous
broadening is small. Under this condition the spin relaxation can be suppressed,
especially for that along the growth direction of the quantum well. 
This cancellation
effect may cause a peak (the cancellation peak) in the density or temperature dependence of
the SRT. In this work, we mainly
investigated the electron density and temperature dependences of the
spin relaxation along the quantum-well growth direction. Besides, we also studied the anisotropic
property of the spin relaxation by varying the spin-polarization direction.

In the electron density dependence of the SRT, two peaks can be
observed under proper gate voltages at low temperature: one is the
well-studied density peak\cite{Jiangbulk,Zhang} (also referred to as the 
Coulomb peak in the present work due to the absence of the impurity) and the other
is the cancellation peak. The location of the cancellation peak can be
tuned by the gate voltage. When the two peaks merge together, the SRT
is markedly prolonged. However, if the gate voltage is large enough
and the cancellation effect is absent, only the Coulomb peak can be
observed. Besides, even under a relatively small gate voltage, with
the increase of temperature, the originally existing cancellation peak
disappears due to the weakening of the cancellation effect and 
consequently only one peak can be observed.

In the temperature dependence of the SRT, the presence of the 
cancellation effect also
depends on the gate voltage. When the cancellation effect is
excluded under a large gate voltage, the Coulomb peak in the
temperature dependence, as revealed previously in the systems where the
cancellation effect is absent,\cite{Zhou,Bronold,Ruan,Zhang,Jiangbulk}
can be observed. When the cancellation effect arises with the decrease of
the gate voltage, it together with the
temperature dependence of the scattering lead to a single peak of the SRT
in the temperature dependence. We further investigated the effect of 
impurities on the temperature dependence of the SRT. Under the large
gate voltage where the cancellation effect is absent, the Coulomb peak
is easily destroyed by impurities. On the contrary, under the small gate
voltage where the cancellation effect is present, a peak always exists
even with very high impurity density. 

We also showed the anisotropy of the spin relaxation with respect to
the spin-polarization direction. It was found that the
cancellation peak gradually disappears when the
spin-polarization direction is tilted from the quantum-well growth
direction to the in-plane one. This is because  the
inhomogeneous broadening contributed by the effective magnetic field
along the growth direction of the quantum well
becomes more important and the cancellation effect becomes weaker.

\begin{acknowledgments}
One of the authors (MWW) would like to thank X. Marie for valuable
discussions.
This work was supported by the Natural Science Foundation of China
under Grant No.\ 10725417, the National Basic Research Program of
China under Grant No.\ 2006CB922005 and the Knowledge Innovation
Project of Chinese Academy of Sciences.
\end{acknowledgments}

\end{document}